\documentclass[prb,reprint,twocolumn,showpacs,superscriptaddress]{revtex4-1}
\usepackage{amsmath, amssymb, amsthm, braket}
\usepackage[breaklinks]{hyperref}
\usepackage{graphicx}
\usepackage{subfigure}
\usepackage{listings}

\usepackage[usenames,dvipsnames]{color}
\definecolor{purple}{rgb}{0.5,0,0.5}

\begin{document}

\title{Designer non-Abelian anyon platforms: from Majorana to Fibonacci}

\author{Jason Alicea}
\affiliation{Department of Physics and Institute for Quantum Information and Matter, California Institute of Technology, Pasadena, California 91125, USA}
\author{Ady Stern}
\affiliation{Department of Condensed Matter Physics, Weizmann Institute of Science, Rehovot, 76100, Israel}

\begin{abstract}

The emergence of non-Abelian anyons from large collections of interacting elementary particles is a conceptually beautiful phenomenon with important ramifications for fault-tolerant quantum computing.  Over the last few decades the field has evolved from a highly theoretical subject to an active experimental area, particularly following proposals for trapping non-Abelian anyons in `engineered' structures built from well-understood components.  In this short overview we briefly tour the impressive progress that has taken place in the quest for the simplest type of non-Abelian anyon---defects binding Majorana zero modes---and then turn to similar strategies for pursuing more exotic excitations.  Specifically, we describe how interfacing simple quantum Hall systems with conventional superconductors yields `parafermionic' generalizations of Majorana modes and even Fibonacci anyons---the latter enabling fully fault tolerant universal quantum computation.  We structure our treatment in a manner that unifies these topics in a coherent way.  The ideas synthesized here spotlight largely uncharted experimental territory in the field of quantum Hall physics that appears ripe for discovery.

\end{abstract}

\maketitle

\section{Introduction}

All fundamental particles invariably conform to the familiar boson-fermion dichotomy, as topology disallows richer exchange statistics for point-like particles moving in continuous 3D space.\footnote{This rather sweeping statement derives from the fact that exchanging two point-like particles twice in continuous 3D space is topologically equivalent to keeping both particles fixed.  Individual exchanges can thus at most introduce a minus sign into the wavefunction since `squaring' the operation must produce the identity.}
A great virtue of condensed matter is that many \emph{interacting} bosons and fermions can nevertheless exhibit collective behavior that sharply defies intuition based on our understanding of the elementary microscopic constituents.  Exotic topological phases of matter supporting anyons---emergent quasiparticles that harbor a novel form of exchange statistics---provide an elegant illustration.  The subject of anyons grew out of the important realization that topology places much less stringent restrictions on the statistics of identical particles in 2D relative to 3D systems.\cite{Leinaas,WilczekAnyons,SternReview,TQCreview}  Conceptually attractive physical realizations are bound states of charge and flux in 2D electron systems; due to the Aharonov-Bohm effect adiabatically braiding these charge-flux composites around one another generically yields a complex statistical phase intermediate between $+1$ and $-1$.  There is little doubt that such `Abelian anyons' (so named because sequential exchanges commute) appear almost universally in fractional quantum Hall systems even though direct experimental confirmation is challenging.

While already quite fascinating, the plot thickens considerably in the case of non-Abelian anyons.  These more exotic (and elusive) quasiparticles carry `internal' zero-energy degrees of freedom that encode a ground-state degeneracy immune to local perturbations.  Braiding non-Abelian anyons produces not just a statistical phase factor, but rather non-commutatively rotates the system within the degenerate manifold.  In other words, the quantum state itself changes despite the indistinguishability of the quasiparticles undergoing the exchange.  This phenomenon of non-Abelian statistics---an undeniably beautiful piece of physics in its own right---may ultimately prove useful as well due to tantalizing topological quantum computing applications.\cite{kitaev,TQCreview,SternLindner}  The key idea is that one can non-locally embed  qubits in the system's degenerate ground-state wavefunctions; moreover, braiding the anyons non-locally manipulates these qubits by virtue of non-Abelian statistics.  Within this approach local environmental perturbations are not removed---they largely become irrelevant.  Put more quantitatively, this approach allows for decoherence times that grow exponentially as temperature decreases, as opposed to the power-law increase that typically characterizes systems without topological protection.

\begin{figure}
\centering
\includegraphics[width=3in]{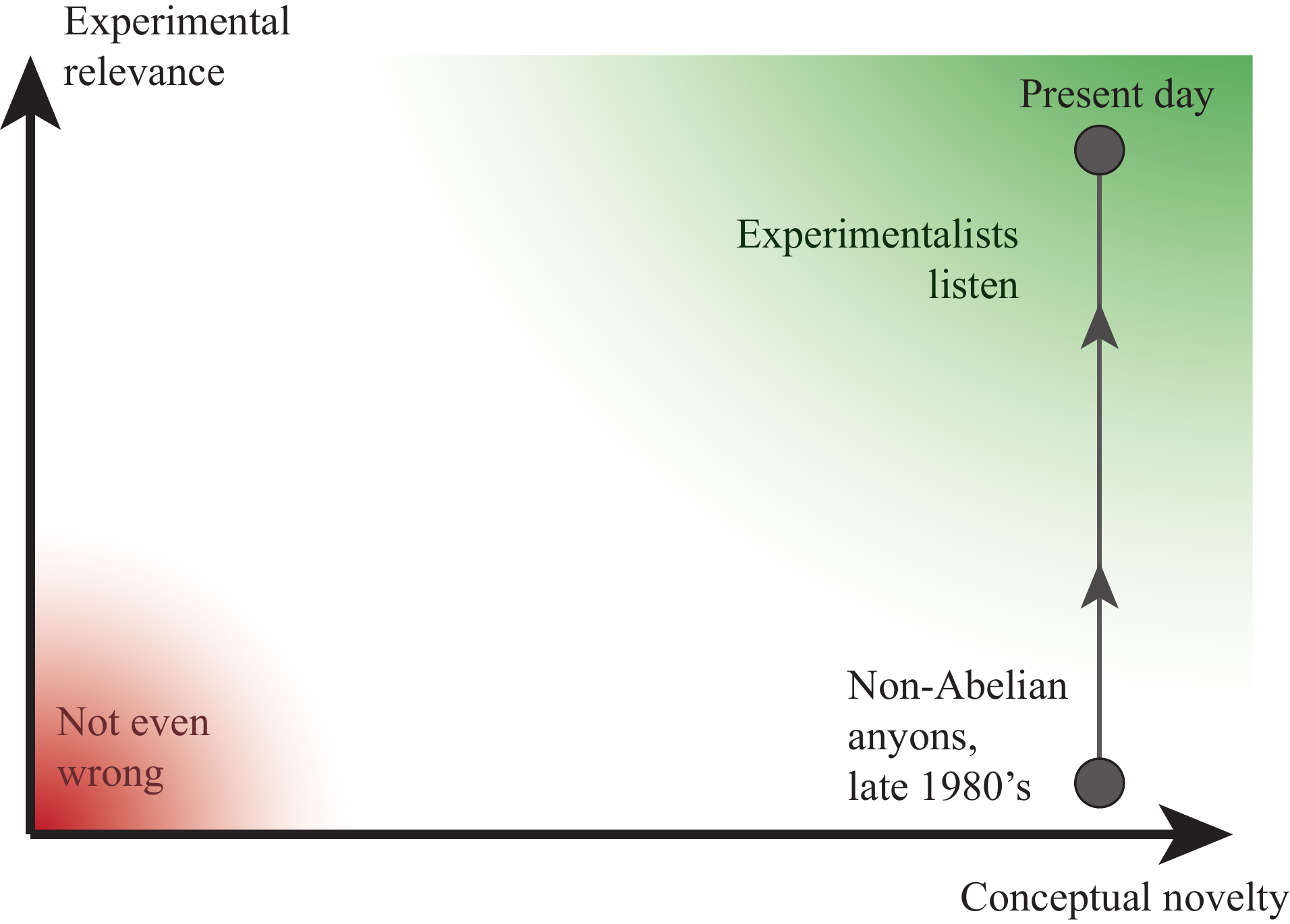}
\caption{Fisher plot quantifying problems in theoretical physics. 
The study of non-Abelian anyons has attained increasing experimental relevance beginning in the early 1990's, with a particularly rapid ascension following the advent of designer non-Abelian platforms.  Note that in the figure we use `listen' to mean `hear out' rather than `obey'.  }
\label{FisherPlot}
\end{figure}

Topological quantum computing supplants the infamous decoherence problem with a quite different challenge: where to find suitable non-Abelian anyons in nature with which to fabricate the hardware?  If easily answered scalable quantum computers might already exist.
During the last three decades the pursuit of non-Abelian anyons has evolved dramatically following a series of pioneering theoretical insights.
It is amusing to view this evolution in terms of the `Fisher plot' (Figure \ref{FisherPlot}) characterizing problems in theoretical physics in terms of their conceptual novelty and experimental relevance.\footnote{This plot was introduced to J.~A.~by Matthew P.~A.~Fisher, hence the name.}  The theoretical inception of non-Abelian anyons as physically permissible objects took place in the late 1980's through impressive (and formidable) mathematical physics work.\cite{MooreSeiberg:89,Witten:JonesPolynomial:1989,Frohlich,Fredenhagen,Froehlich}  At this early stage the problem would surely have resided far along the horizontal axis of Fig.~\ref{FisherPlot}, as non-Abelian anyons revealed an absolutely spectacular facet of quantum mechanics but with no obvious home.

The first of many major boosts towards experimental relevance transpired in 1991 with Moore and Read's introduction of a candidate non-Abelian quantum Hall phase mysteriously dubbed the Moore-Read state.\cite{MooreRead}
Almost ten years later Read and Green observed that the non-Abelian physics of the Moore-Read state---a highly nontrivial strongly correlated phase---could be emulated in a weakly correlated spinless 2D $p+ip$ superconductor.\cite{ReadGreen}  Note that the latter shares many properties with He-3 studied to great effect by Volovik and others.\cite{VolovikBook}  The non-Abelian anyons found in the Moore-Read phase are replaced by order-parameter defects in the superconductor, namely $h/2e$ vortices that bind Majorana zero modes.  
While spinless 2D $p+ip$ superconductivity is readily understood theoretically, finding experimental examples poses greater difficulty as electrons stubbornly carry spin, usually roam in 3D, and almost always prefer to pair in the $s$-wave channel.  Fortunately, in an influential 2008 work Fu and Kane devised a way of engineering the physics of spinless 2D $p+ip$ superconductivity by interfacing 3D topological insulators with \emph{conventional} superconductors.\cite{FuKane}  Their proposed `designer' non-Abelian anyon platform, which laid the groundwork for a now extensive literature of similar ilk, takes serendipity out of the equation; the electrons are given an offer they can't refuse.

Experimental prospects brightened further still with the realization that non-Abelian anyons are not actually unique to the 2D world.  One can `cheat' the constraints of topology in one of two ways: $(i)$ by utilizing particles that are not point-like\cite{TeoKane,Freedman3D,Freedman3Db,LevinLoops,WenLoops,WenLoops2,QiLoops} or $(ii)$ by restricting particles to move along discrete 1D paths in spaces of arbitrary dimension\cite{HalperinBraiding}.
The latter loophole allows one to harness non-Abelian statistics by piecing together 1D systems to form networks.\cite{AliceaBraiding,ClarkeBraiding,HalperinBraiding,BondersonBraiding,BeenakkerBraiding,BeenakkerBraiding2,Necklace}
Particularly appealing building blocks are 1D spinless $p$-wave superconductors that also harbor Majorana zero modes\cite{1DwiresKitaev} (at domain walls rather than vortices); these phases can be similarly engineered using existing materials, but in arguably simpler geometries compared to their 2D counterparts\cite{MajoranaQSHedge,1DwiresLutchyn,1DwiresOreg}.  We will encounter other more exotic examples later on.

Following the distillation of non-Abelian platforms into ever-simpler forms, the problem now enjoys a great deal of experimental activity and unequivocally belongs in the upper-right corner of Fig.~\ref{FisherPlot}.  In this article we provide a lighthearted overview of some recent developments in this ongoing distillation process.  We begin in Sec.~\ref{Majorana} by briefly reviewing the basic concepts underlying designer 1D and 2D superconducting Majorana platforms and then highlight select experiments that they have inspired\cite{mourik12,das12,Rokhinson,deng12,finck12,Churchill,chang12,Kurter,Orlyanchik,Shabani} (many of which are already bearing fruit).  For a more complete discussion we refer readers to the many excellent Majorana reviews in Refs.~\onlinecite{BeenakkerReview,FlensbergReview,TewariReview,FranzReview,BeenakkerReview2} as well as the reviews of Refs.~\onlinecite{AliceaReview,AliceaNN}.

Encouraged by preliminary successes in the Majorana quest, the remaining sections explore similar strategies to design more exotic types of non-Abelian anyons.  Along the way we will necessarily forgo the comforts of non-interacting electron physics and re-enter the strongly correlated realm.  Crucially, however, with an engineering spirit we will build up the physics using only presently available, well-understood phases of matter: ordinary superconductors and conventional fractional quantum Hall phases.  In our view this route is both intrinsically interesting and strongly motivated by quantum computing applications.  Indeed, even with perfect control over systems supporting Majorana modes, braiding alone enables rather limited fault-tolerant quantum information processing.  Section \ref{parafermions} surveys recent proposals for stabilizing `parafermionic' generalizations\cite{Fendley} of Majorana zero modes.  These, too, turn out to fall short of allowing computational universality.   We will describe in Sec.~\ref{Fibonacci} that parafermion zero modes can, however, serve as building blocks for a new fractionalized superconductor supporting so-called Fibonacci anyons---which \emph{do} permit fully universal decoherence-free quantum computation.  The underlying physics leading to Fibonacci anyons (forcing bosons built from fractionalized degrees of freedom to condense) suggests a general strategy towards realistic topological quantum computing hardware that warrants further theoretical attention.

\section{Majorana materializes}
\label{Majorana}

\begin{figure}
\centering
\includegraphics[width=\columnwidth]{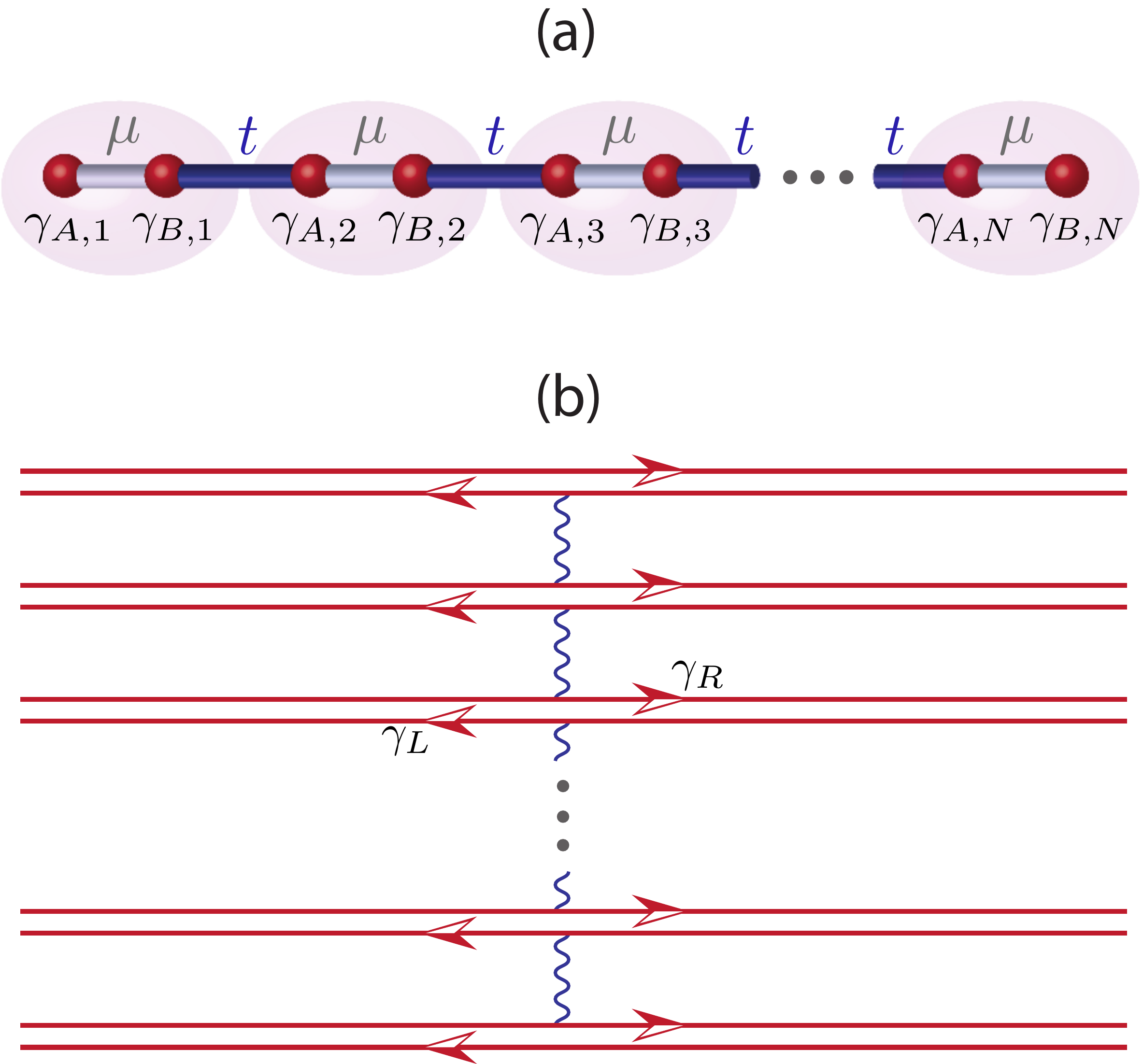}
\caption{(a) Kitaev chain in the limit $t = \Delta$.   (b) Construction of a topological 2D spinless $p+ip$ superconductor from critical Kitaev chains that host chiral Majorana modes $\gamma_R$ and $\gamma_L$.}
\label{MajoranaFig}
\end{figure}

The `Kitaev chain'\cite{1DwiresKitaev} carries immense pedagogical value as a minimalist caricature of a 1D spinless $p$-wave superconductor.  
The Hamiltonian reads
\begin{equation}
  H = \sum_x\left[-\mu c_x^\dagger c_x - \frac{1}{2}\left(t c_x^\dagger c_{x+1} + \Delta c_x c_{x+1} + {\rm H.c.}\right) \right],
\end{equation}
where $c_x$ represents a spinless fermion at site $x$ with chemical potential $\mu$, hopping amplitude $t$, and $p$-wave pairing strength $\Delta$.  One can revealingly recast the model in terms of Majorana fermion operators $\gamma_{A/B,x}$ (which are Hermitian and square to the identity) by writing $c_x = (\gamma_{B,x}+i \gamma_{A,x})/2$.  In the limit $t = \Delta$ the Hamiltonian then nicely simplifies to
\begin{equation}
  H = -\frac{i}{2}\sum_{x} (-\mu\gamma_{A,x}\gamma_{B,x} + t \gamma_{B,x}\gamma_{A,x+1}).
  \label{Hmajorana}
\end{equation}
As Fig.~\ref{MajoranaFig}(a) illustrates the first and second terms favor competing dimerizations for the Majorana operators.  When $|\mu| > t$ the former wins and the chain realizes a gapped superconductor that smoothly connects to a trivial product state upon sending $t \rightarrow 0$.  At $|\mu| = t$ the terms compete to a draw, yielding interesting critical behavior that we elaborate on shortly.  Most remarkably, a gapped topological superconducting phase appears when $|\mu| < t$.  Here `unpaired' Majorana zero modes bind to the ends of the chain, which Fig.~\ref{MajoranaFig}(a) makes visually obvious in the extreme $\mu = 0$ limit.  These modes commute with the Hamiltonian but not with each other; a ground-state degeneracy must therefore exist.

More physically, the zero modes encode a two-fold ground-state degeneracy consisting of states with even and odd total fermion number---a striking feature given that superconductors typically prefer even parity so that all electrons simultaneously Cooper pair.  One could of course envision ordinary superconductors with `accidental' zero-energy bound states that also produce a degeneracy between even- and odd-parity states.  However, the latter degeneracy requires fine tuning, and moreover a local measurement of the bound state's occupation number distinguishes the ground states.  In sharp contrast, Majorana zero modes in the Kitaev chain form a single fermionic state fragmented across arbitrary distances.  The associated ground-state degeneracy thus remains exact in the thermodynamic limit and requires absolutely no fine tuning.  As a corollary no local measurement can detect, even in principle, which ground state the system realizes.

One can further leverage the Kitaev chain to access the topological phase of a 2D spinless $p+ip$ superconductor---which harbors a gapless chiral Majorana mode at its boundary and (as a direct consequence\footnote{One can view a vortex core as a puncture in the superconductor that generates a new chiral Majorana edge state.  With even vorticity the edge spectrum yields only modes with non-zero energy due to finite-size quantization; odd vorticity, however, shifts the spectrum in such a way that a single Majorana zero mode appears at the vortex core. The total number of Majorana zero modes, bound to vortices and on the sample's edge, must always be even.}) Majorana zero modes at $h/(2e)$ vortex cores\cite{ReadGreen}.  Consider a 2D system formed by initially decoupled chains described by Eq.~\eqref{Hmajorana}.  With $\mu = -t$ each chain realizes a gapless (Ising\footnote{Equation \eqref{Hmajorana} indeed maps precisely to the transverse-field Ising model under a Jordan-Wigner transformation.  Note that we have now used the word `Ising' on multiple occasions, successfully avoiding the omission correctly noted in footnote 1 of Ref.~\onlinecite{Fendley}.}) critical point separating topologically trivial and non-trivial phases as noted above.  Here right- and left-moving Majorana fields $\gamma_{R/L}$, with velocity $v\propto t$, capture the low-energy physics.  Indeed, upon expanding $\gamma_{A,x} \sim \gamma_R + \gamma_L$ and $\gamma_{B,x} \sim \gamma_R - \gamma_L$ the critical Hamiltonian for each chain takes the form
\begin{equation}
  H_{\rm crit} = \int_x\left[-i\hbar v\left(\gamma_R \partial_x\gamma_R -\gamma_L\partial_x\gamma_L\right)\right].
\end{equation}
Successively coupling the right-mover from one chain with a neighbor's left-mover drives the system into the topological phase of a 2D spinless $p+ip$ superconductor.  Figure \ref{MajoranaFig}(b) shows that the telltale `unpaired' chiral Majorana modes appear at the upper and lower boundaries.

The notion of \emph{central charge} from conformal field theory implicitly plays an important role in the preceding discussion.  Central charge $c$ essentially counts the number of gapless modes in a 1D system and intimately relates to heat conductance.  An ordinary fermionic mode, for instance, carries $c = 1$; a gapless Majorana mode---which represents `half' of an ordinary fermion---correspondingly carries $c = 1/2$.  One can thus view the Kitaev chain at the critical point as a partially gapped spinless wire in which a pair of gapless Majorana modes take the place of two $c = 1$ modes.  Hybridizing the remnants of these $c = 1$ modes with adjacent chains then illuminates a pathway into nontrivial higher-dimensional phases whose existence may be relatively obscure from the original fermionic degrees of freedom.
Bootstrapping off of 1D chains to controllably access exotic 2D physics constitutes a powerful theoretical tool that we encounter again in Sec.~\ref{Fibonacci}.

The minimalist caricatures above serve as inspiration for more realistic portraits of 1D and 2D topological superconducting phases supporting Majorana modes.  Devising plausible experimental blueprints requires effectively stripping the spin from electrons in reduced-dimensional structures, and then somehow forcing those `spinless' fermions to Cooper pair.  In recent years theorists have proposed a myriad of clever schemes for overcoming these challenges; for a partial list see Refs.~\onlinecite{BeenakkerReview,AliceaReview}.  Nearly all follow the same unifying recipe, which for clarity we outline in the context of 1D topological superconductors:

$(i)$ A garden-variety 1D electron system typically contains two right-moving and two left-moving electron modes due to spin degeneracy.  Remove half of these modes so that at low energies the problem maps to spinless fermions.  One can achieve such a band structure in exactly two ways.  Either \emph{spatially} separate the two sets of right- and left-movers, as effectively occurs at the edge of a 2D quantum spin Hall system\cite{KaneMele}, or break time-reversal symmetry to \emph{energetically} separate the bands and eliminate one pair of modes.

$(ii)$ Impose nontrivial spin structure among the remaining right-mover and left-mover.  That is, spin-orbit interactions of some form must ensure that electrons at the Fermi energy do not perfectly spin-polarize.  Note that strong spin-orbit coupling appears by definition in quantum spin Hall edges but must be incorporated explicitly in other 1D settings.

$(iii)$ Finally, couple the `spinless' 1D fermions to a conventional $s$-wave superconductor.  By virtue of the proximity effect, the 1D modes of interest inherit spin-singlet Cooper pairing---which efficiently forms since both spin up \emph{and} spin down states remain available at low energies.  A 1D topological superconductor awaits, though for best results cooling is strongly recommended.

With only minor straightforward modifications the above recipe generalizes to 2D topological superconducting platforms as well.

Such schemes outline surprisingly simple pathways towards remarkably exotic physics.  Consistent with Fig.~\ref{FisherPlot}, the experimental community has in kind responded quite favorably.  The groundbreaking work of Lutchyn et al.\cite{1DwiresLutchyn} and Oreg et al.\cite{1DwiresOreg} on semiconducting-wire-based setups inspired a wave of experiments\cite{mourik12,das12,Rokhinson,deng12,finck12,Churchill,chang12} that reveal tantalizing evidence of Majorana zero modes.  Experiments on these systems continue with steady, impressive progress today.  Chains formed by magnetic atoms on a superconductor provide another fruitful venue for Majorana physics.  While initial theories invoked spiral magnetic order as a key component for topological superconductivity\cite{NoSOC1,Yazdani,KlinovajaRKKY,Braunecker,Reis,KimMagneticAtoms}, recent experiments\cite{YazdaniTalk} suggest that a more likely mechanism may arise from \emph{ferromagnetic} moments hybridizing with a strongly spin-orbit-coupled superconductor (similar to semiconducting-wire platforms\cite{1DwiresLutchyn,1DwiresOreg}).  The original Fu-Kane proposals for topological superconductivity in quantum spin Hall edges\cite{MajoranaQSHedge} and 3D topological insulator surfaces\cite{FuKane} of course also remain influential.  In particular, the former setting appears increasingly promising due to works that successfully induce proximity effects in HgTe\cite{HgTeProximity} and InAs/GaSb\cite{KnezProximity,Pribiag} quantum wells; in parallel experiments on 3D topological insulators show similar encouraging progress (see, e.g., Refs.~\onlinecite{Kurter,Orlyanchik}).  Interpreting this body of experiments poses a subtle yet fascinating problem that we will not dive into here.  

This state of affairs is truly exciting---the field steadily marches along complementary pathways towards the lofty goal of definitively observing and manipulating Majorana zero modes.  The main desired manipulation is braiding to establish non-Abelian statistics in the laboratory.  In 2D braiding is rather easy to grasp and straightforward to implement (mentally).  Here Majorana modes localize at points in the plane, e.g., at vortex cores.  A braid proceeds by adiabatically evolving their positions until all Majoranas return to their initial locations, up to permutations.  Each braid corresponds to a unitary transformation that acts within the ground-state subspace and specifies how the system's quantum state rotates under the exchange. Crucially, the results depend only on the braid topology \emph{provided} the ground-state degeneracy does not change during the process.  Hence we require that the Majorana zero modes remain well-separated at every step of the exchange since tunneling between adjacent pairs generically splits the degeneracy.  The set of unitary braid transformations then forms a non-Abelian representation of the braid group.

Extending the notion of braiding to 1D systems poses greater difficulty, as the reader may appreciate by attempting to (lawfully!)~interchange two cars along a narrow one-lane street.  The most conceptually straightforward way out involves patterning these 1D systems to mimic a traffic grid.  Majorana zero modes---like the vehicles---can adiabatically exchange positions with the aid of an `alleyway', all the while maintaining the separation required to preserve the ground-state degeneracy. The alleyway releases the system from the confines of 1D in a minimal way that permits the interchange.

Alternatively, braiding can arise from a closed trajectory in parameter space rather than real space. Consider a Hamiltonian that supports a set of Majorana zero modes and depends on some externally controllable parameters.  Cycling the Hamiltonian parameters along a closed trajectory can \emph{effectively} generate a braid even if the defects binding the zero modes remain static or move only within one dimension.\footnote{Such an approach is familiar from `measurement-only' topological quantum computation schemes\cite{MeasurementOnlyTQC}, though does not require measurements.}
Continuing with the one-lane street metaphor, quantum mechanics offers the frustrated drivers the enticing possibility to interchange their cars through tunneling.
Indeed, by varying tunnel couplings the drivers can \footnote{Patent pending.} create an empty parking spot along the lane and `teleport' the first car to that spot. The second car may then teleport into the first's initial position, followed by the first car teleporting again to complete the exchange. All these steps---the creation of the empty parking spot and the sequential `teleportations'---may occur while keeping the ground-state degeneracy fixed.\footnote{Unfortunately, however, these ideas appear impractical for improving the situation on the busy Los Angeles and Tel Aviv parking lanes.}

The unitary transformation that results from this quantum version of urban life precisely matches that arising from the leisurely braiding of Majorana modes in 2D.\cite{AliceaBraiding}
These transformations, while enjoying topological protection, provide only one of the three elementary gates required for a universal gate set.  Thus one can not approximate arbitrary qubit rotations solely through braiding of defects binding Majorana zero modes.  
A natural question therefore arises: Can we move further along the horizontal axis in Fig.~\ref{FisherPlot} in search of experimental blueprints for still more exotic anyons with greater utility for quantum computation?  This is the subject to which we turn next.

\section{Beyond Majorana: zero modes from (simple) fractional quantum Hall edges}
\label{parafermions}

In 2012 Fendley\cite{Fendley} generalized the Kitaev chain in an interesting way that inspired numerous proposals for moving beyond the Majorana paradigm.  Fendley's model replaces the Majorana operators in Fig.~\ref{MajoranaFig}(a) with unitary $\mathbb{Z}_M$ \emph{parafermion} operators $\alpha_j$, where $j$ labels sites of the chain.  These operators satisfy
\begin{equation}
  \alpha_j^M = 1,~~~~~~~~\alpha_j \alpha_{j'>j} = e^{i 2\pi/M} \alpha_{j'}\alpha_j
  \label{ParafermionOps}
\end{equation}
for some integer $M> 2$.  Notice that setting $M = 2$ recovers the familiar Majorana algebra.  Precisely as in the Kitaev chain the bonds indicated by $\mu$ and $t$ in Fig.~\ref{MajoranaFig}(a) favor competing parafermion dimerizations.  If the former wins a trivial gapped state emerges; if the latter dominates the system enters a nontrivial topological phase with `unpaired' parafermionic zero modes that encode an $M$-fold ground-state degeneracy.  (Note that various nomenclature for these modes appears in the literature, including generalized Majorana modes, fractionalized Majoranas, and sometimes parafendleyons.)  Very interesting critical behavior can intervene between these phases but we defer a discussion to the next section.

Immediate obstacles arise when seeking designer platforms for parafermion zero modes.  Most glaringly, in contrast to the Kitaev chain the operators in Fendley's toy model are neither bosonic nor fermionic, which greatly obscures candidate host systems.  A subtler problem lurks as well.  Rigorous classifications of gapped phases for generic 1D systems---including with strong interactions---allow for Majorana zero modes but, sadly, nothing more exotic\cite{FidkowskiKitaev1,FidkowskiKitaev2,Turner,ChenGuWen,Schuch} (at least without special symmetries). The \emph{anyonic} commutation relations in Eq.~\eqref{ParafermionOps}, however, hint at how one can escape both difficulties.  Specifically, these same commutation relations appear for operators creating fractionalized quasiparticles along edges of `simple' fractional quantum Hall systems that occur in many materials and whose physics is well understood.  This intuition (correctly) suggests that we can engineer parafermion zero modes in `wires' assembled from quantum Hall edges---crucially, without betraying our engineering principles.  Such `wires', which are \emph{not} strictly 1D systems, do not fall victim to the constraints dictated by 1D classifications.

As a primer it is instructive to revisit Fu and Kane's proposal for trapping Majorana zero modes on a 2D topological insulator edge.\cite{MajoranaQSHedge}  The edge hosts a single set of counterpropagating modes that can gap out via two `incompatible' mechanisms: $(i)$ incorporating magnetism to backscatter the edge electrons or $(ii)$ Cooper pairing right- and left-movers  (\`{a} la the recipe from Sec.~\ref{Majorana}).  As one would naturally expect it is not possible to evolve smoothly from one kind of gap to the other without crossing a phase transition.  Domain walls separating spatial regions gapped by these different means thus feature interesting physics; in this case such defects bind localized Majorana zero modes.

\begin{figure}
\centering
\includegraphics[width=\columnwidth]{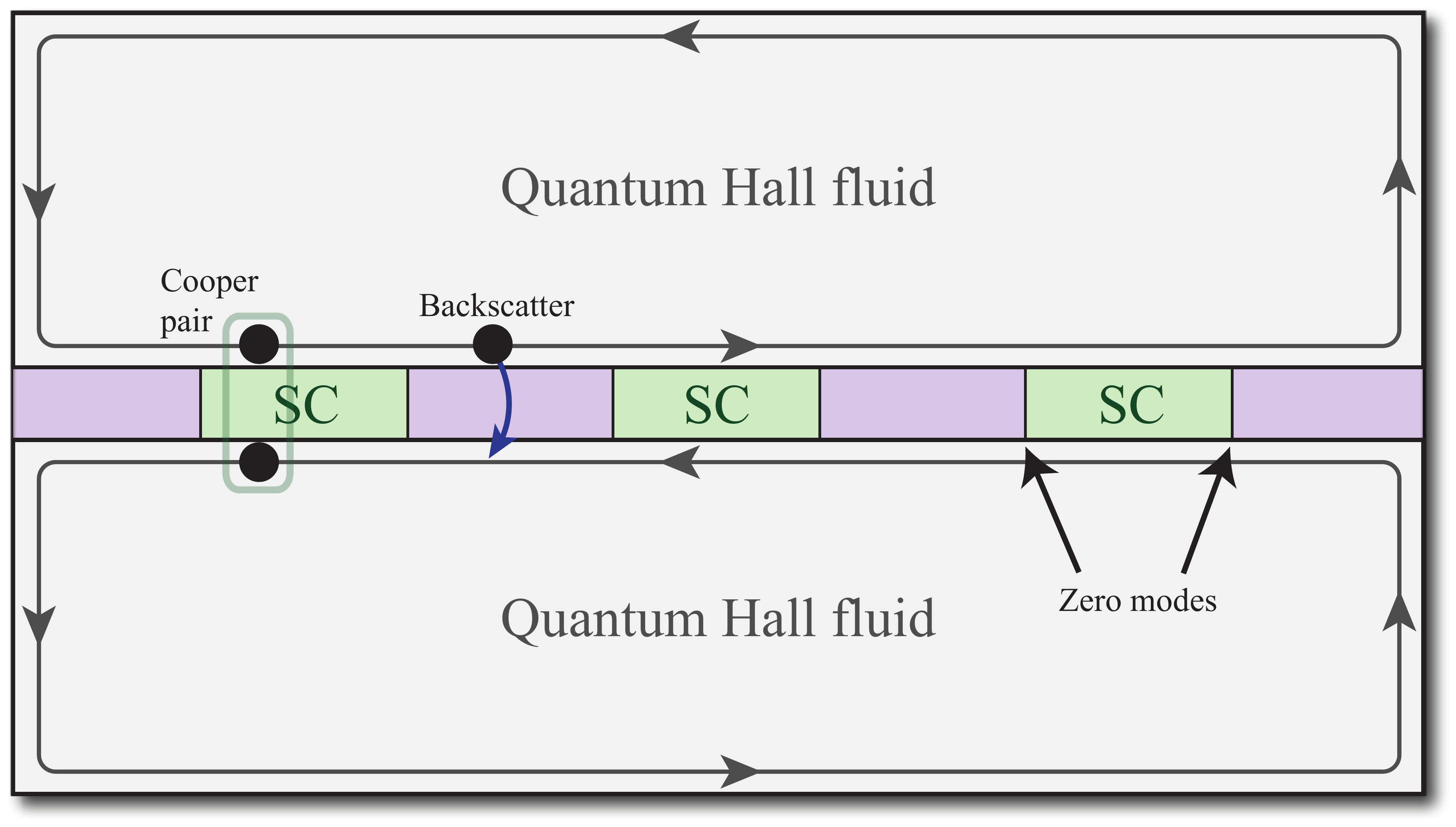}
\caption{1D `wire' formed by counterpropagating quantum Hall edge states.  Regions labeled `SC' represent superconductors.  Domain walls between segments gapped by backscattering and Cooper pairing bind Majorana zero modes (in the integer case) and parafermionic generalizations (in the fractional case).  }
\label{QHfig}
\end{figure}

We can mimic the same physics in integer quantum Hall systems by arranging two filling factor $\nu = 1$ fluids side by side.  As Fig.~\ref{QHfig} illustrates their interface supports counterpropagating modes---just like the topological insulator edge---that can again acquire a gap through either backscattering, or Cooper pairing induced by a proximity effect.  (To generate pairing here one could for instance employ a parent superconductor with a triplet component; we describe variations below, however, that utilize simple $s$-wave superconductors.)  Majorana modes localize to domain-wall defects here too.  Yet another realization is a bilayer in which one layer carries electrons at $\nu=1$ while the other has $\nu = -1$ holes---same magnetic field and density but  opposite charge and possibly also opposite spins.  In any of these setups each superconducting-gapped region can accommodate an extra electron for free.  This is the ground-state degeneracy encoded by the Majorana zero modes.

Consider next Fig.~\ref{QHfig} fabricated with \emph{fractional} quantum Hall fluids realizing $\nu = 1/3$ Laughlin states.  The key difference from the integer case is that our `wire' built from counterpropagating edge states now accommodates fractionalized charge-$e/3$ excitations---leading to richer defect physics.  Consistent with the intuition laid out above, domain walls separating pairing- and backscattering-gapped regions correspondingly trap $\mathbb{Z}_6$ \emph{parafermion} zero modes satisfying Eq.~\eqref{ParafermionOps} with $M = 6$.\cite{LindnerParafendleyons,ClarkeParafendleyons,ChengParafendleyons}  These modes commute with the Hamiltonian\footnote{Technically exact commutation with the Hamiltonian only arises in an extreme limit; see Refs.~\onlinecite{Fendley,Jermyn}.  The ground-state degeneracy, however, survives even away from this extreme limit.} but not with one another, implying a ground-state degeneracy like in the Majorana problem.

The ground-state degeneracy guaranteed by the parafermion zero modes admits a clear physical interpretation.  One may profitably view the interface in Fig.~\ref{QHfig} as an array of superconducting quantum dots immersed in fractional quantum Hall fluid.  Importantly, the charge of each dot is quantized (modulo a Cooper pair) in units of the elementary charge for the surrounding insulating regions---in this case $e/3$.  Thus the system's states are characterized by a set of charge quantum numbers for the dots that take one of six values $0, 1/3, 2/3, 1, 4/3, 5/3$ and sum to an integer.  The special feature of this setup is that \emph{all} such charge configurations yield the same energy.  In other words each superconducting dot can, remarkably, absorb a \emph{fractional} charge of $e/3$ without energy penalty!  The `quantum dimension' of the domain walls---which counts the asymptotic number of ground states per defect---thus equals $\sqrt{6}$.  Majorana defects, by contrast, yield an exponentially smaller degeneracy and carry a quantum dimension of $\sqrt{2}$.

In practical terms, the experimental setting introduced above is certainly much more demanding than that needed for stabilizing Majorana modes.  For one Laughlin states are spin-polarized---thus increasing the difficulty of Cooper-pair formation in a single layer.  Moreover, the combination of superconductivity and high magnetic fields required for the quantum Hall effect poses a highly nontrivial challenge.  One can alleviate the first issue by replacing the Laughlin states with spin-unpolarized $\nu = 2/3$ fluids.\cite{Mong,ClarkeCircuits}  The `wire' in Fig.~\ref{QHfig} then carries additional structure since each edge contributes a charge mode \emph{and} a backwards-propagating neutral mode.  Domain walls between regions gapped by spin-conserving electron backscattering and $s$-wave Cooper pairing bind $\mathbb{Z}_3$ parafermion zero modes obeying Eq.~\eqref{ParafermionOps} with $M = 3$ rather than 6.  Each superconducting dot can then freely accommodate charge $2e/3$.  This case captures the minimal generalization of Majorana defects and plays a special role in the next section.

Other variations are also possible---including quantum Hall setups that do not invoke superconductivity at all.\cite{ChernInsulatorParafendleyons,BarkeshliParafendleyons1,BarkeshliParafendleyons2,BarkeshliClassification1,BarkeshliClassification2}  Very generally, stabilizing zero modes simply requires incompatible means of gapping out an edge-state `wire'.  If a quantum Hall edge supports more than one mode (as in all fillings not of the Laughlin series), there may exist multiple \emph{charge-conserving} mechanisms for opening a gap.  Domain-wall defects can then similarly trap Majorana or parafermion zero modes depending on details.  Superconductivity plays an illuminating but ultimately inessential part of the story, though alternative methods carry their own challenges.  For related studies, including proposals for zero-mode detection, see Refs.~\onlinecite{VaeziParafendleyons,Hastings,QuantumWiresParafendleyons,ParafendleyonLattice,Klinovaja1,Klinovaja2,BarkeshliDetection,Klinovaja3,ClarkeCircuits,ZhangKane,Orth,Li,Tsvelik}.

The prescriptions for braiding Majorana defects in 1D settings reviewed in Sec.~\ref{Majorana} also allow, albeit with some care, interchange of parafermion zero modes in the quantum Hall architectures surveyed above.\cite{LindnerParafendleyons,ClarkeParafendleyons}  As one may expect, denser unitary transformations arise relative to the Majorana case.  The added richness, however, remains insufficient to allow universal quantum computation as braiding yields only two of the three elementary gates required for universality.  To seek hardware for a fully fault-tolerant topological quantum computer we must move still further along the horizontal axis in Fig.~\ref{FisherPlot}.

\section{Towards universal topological quantum computation}
\label{Fibonacci}

In the long-term quest for a universal topological quantum computer so-called `Fibonacci anyons' constitute one of the holy grails.
Contrary to the non-Abelian defects discussed so far, a single gate obtained by a clockwise braid of these quasiparticles suffices to approximate \emph{arbitrary} qubit rotations within any desired accuracy.\cite{TQCreview,Freedman02a,Freedman02b}  Fibonacci anyons obey the peculiar property that a pair can either annihilate one another or beget a new Fibonacci anyon.  Their name derives from the fact that as one nucleates additional anyons the ground-state degeneracy of the host system grows as the Fibonacci sequence; their quantum dimension therefore equals the golden ratio, $(1+\sqrt{5})/2$.

\begin{figure}
\centering
\includegraphics[width=\columnwidth]{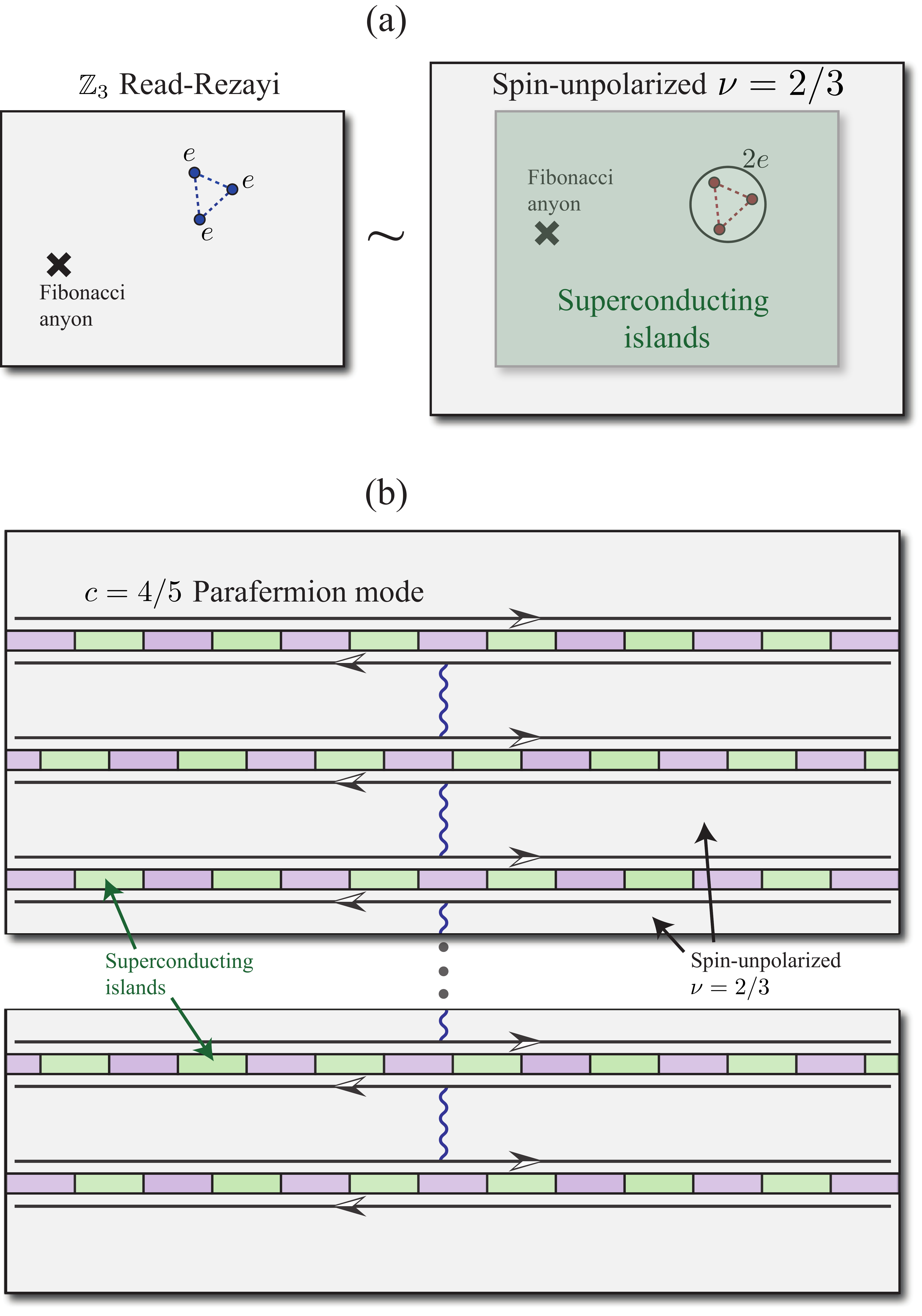}
\caption{(a) Correspondence between the $\mathbb{Z}_3$ Read-Rezayi quantum Hall state and a $\nu = 2/3$ quantum Hall/superconductor heterostructure.  Both harbor Fibonacci anyons that enable universal topological quantum computation.  (b) Array of gapless parafermion modes that emerge upon partially gapping the edge-state `wire' from Fig.~\ref{QHfig}.  Hybridizing these modes via the wavy lines accesses the phase with Fibonacci anyons from a weakly coupled chain perspective.  The philosophy is identical to that of Fig.~\ref{MajoranaFig}(b) for the Majorana case.  }
\label{FibFigure}
\end{figure}

Nature may provide an `intrinsic' source of Fibonacci anyons in certain highly exotic non-Abelian quantum Hall fluids.  In 1999 Read and Rezayi introduced a quantum Hall series that generalizes the famed Moore-Read state.\cite{ReadRezayi}  The first member of this series---the `$\mathbb{Z}_3$ Read-Rezayi state'---exhibits a particularly interesting anatomy.  Notably, this phase builds in multi-electron clustering correlations.  That is, rather than pairing (as in a superconductor) \emph{triplets} of electrons cluster as sketched in Fig.~\ref{FibFigure}(a).  For various perspectives on this property see Refs.~\onlinecite{ReadRezayi,FendleyFisherNayak2,SternReview}.  Multi-particle clustering underlies fascinating bulk and edge physics.  Aside from the usual charge mode, the boundary of the $\mathbb{Z}_3$ Read-Rezayi state supports a gapless chiral edge state described by a `parafermion conformal field theory' with central charge $c = 4/5$.  (As the names suggest the parafermion \emph{operators} discussed in Sec.~\ref{parafermions} and the \emph{fields} in this conformal field theory relate in a precise way; see Ref.~\onlinecite{LatticeCFTrelation}.)  We can view this $c = 4/5$ sector as an ordinary chiral fermion partially gapped in a subtler fashion compared to the $c = 1/2$ Majorana modes encountered in Sec.~\ref{Majorana}.  As an intimately related consequence the bulk hosts Fibonacci anyons.  The observed quantum Hall state at $\nu = 12/5$ in GaAs quantum wells may actually harbor such a phase, though at present little is known about the observed plateau.  

Can we alternatively \emph{engineer} $\mathbb{Z}_3$ Read-Rezayi physics using phases of matter that we know are currently available in the laboratory?  Doing so requires creating a neutral mode with a fractional central charge on the edge and controllably emulating multi-particle clustering in the bulk of a solid-state device---a highly nontrivial task.  By employing a certain large-$N$ limit\footnote{$N$ denotes the number of coauthors.} Ref.~\onlinecite{Mong} nevertheless successfully introduced a designer \emph{superconducting} Fibonacci anyon platform closely related to the Read-Rezayi phase.  The blueprint, sketched on the right side of Fig.~\ref{FibFigure}(a), involves components familiar from the previous section: a spin-unpolarized $\nu = 2/3$ system coupled to a \emph{two-dimensional} array of conventional superconducting islands.  

We first heuristically motivate why one can naturally expect such a quantum Hall/superconductor hybrid to harbor Fibonacci anyons.  The superconducting islands force charge-$2e$ Cooper pairing into the adjacent $\nu = 2/3$ fluid.  Crucially, however, that medium supports fractionalized excitations, so one should not view the pairs as built merely out of two ordinary electrons.  Rather, the induced Cooper pairs possess finer structure---each arising from triplets of charge-$2e/3$ excitations [see Fig.~\ref{FibFigure}(a)].  Thus the heterostructure leverages pairing, which we can easily obtain, to impose multi-particle clustering analogous to that found in the Read-Rezayi state!

To back up the intuitive picture painted above we can follow a similar algorithm to how we accessed the nontrivial phase of a spinless 2D $p+ip$ superconductor from Kitaev chains in Sec.~\ref{Majorana}.  There we began from decoupled 1D chains and exactly balanced the competing dimerizations illustrated in Fig.~\ref{MajoranaFig}(a).  By doing so we effectively obtained an array of partially gapped wires in which a pair of counterpropagating Majorana modes with central charge $c = 1/2$ describe the low-energy physics.  Hybridizing adjacent critical chains as shown in Fig.~\ref{MajoranaFig}(b) drove the system into a 2D topological superconductor carrying a single chiral Majorana edge state.  We will reproduce precisely this logic to access a Read-Rezayi-like state, but bootstrapping from chains of coupled \emph{parafermion} rather than Majorana modes.\footnote{The beautiful theoretical work by Teo and Kane in Ref.~\onlinecite{TeoKaneChains} originally inspired this strategy.}

With this strategy in mind, consider the setup of Fig.~\ref{FibFigure}(b) in which spin-unpolarized $\nu = 2/3$ quantum Hall fluids generate multiple edge-state `wires' of the type analyzed in Sec.~\ref{parafermions}.  Suppose that the `wires' initially decouple.  One can then view the system as a fractional quantum Hall plane dissected to produce parallel trenches that carry counterpropagating modes with central charge $c = 2$ in each direction.   Following the Majorana example we would like to first partially gap each trench by accessing a critical point featuring a pair of counterpropagating $c = 4/5$ modes.  (Recall that a \emph{chiral} $c = 4/5$ sector lives at the Read-Rezayi edge.)  To do so we alternately gap each trench by Cooper pairing and backscattering, yielding decoupled chains of $\mathbb{Z}_3$ parafermion operators localized to the domain walls.  See Fig.~\ref{FibFigure}(b).   Within each chain two possible dimerizations compete and favor distinct gapped phases---one trivial, the other exhibiting `unpaired' zero modes at the ends.  When these dimerizations precisely balance, a nontrivial critical point emerges described by the non-chiral $c = 4/5$ parafermion conformal field theory that we seek.\footnote{The same theory describes the critical point of the three-state Potts model.  This is not coincidental---the $\mathbb{Z}_3$ parafermion chain maps to precisely that model under a non-local Jordan-Wigner-like transformation.\cite{Fendley,FradkinKadanoff}}  
Tuning to this critical point completes the first step of our algorithm.\footnote{One can alternatively obtain the $c = 4/5$ critical point by adding \emph{spatially uniform} Cooper pairing and backscattering terms that balance one another; for the theory behind this mechanism see the important study by Lecheminant, Gogolin and Nersesyan, Ref.~\onlinecite{LGN}.}  

The second and final step proceeds by judiciously coupling neighboring trenches.  In particular, hybridizing adjacent counterpropagating modes as indicated by wavy lines in Fig.~\ref{FibFigure}(b) fully gaps the bulk leaving an `unpaired' chiral $c = 4/5$ edge state---precisely as in the Read-Rezayi phase.  Physically, the hybridization arises from tunneling of fractional charges through the intervening quantum Hall fluids.\cite{Mong,LatticeCFTrelation}  The resulting `Fibonacci phase' is a superconducting cousin of the $\mathbb{Z}_3$ Read-Rezayi state that, as the edge structure implies, hosts Fibonacci anyons.

The coupled-chain approach followed above provides an illuminating, analytically controllable window into this highly nontrivial 2D physics.  We emphasize though that the Fibonacci phase actually survives well away from the weakly coupled chain limit.  Recent density matrix renormalization group work addresses the broader phase diagram of $\mathbb{Z}_3$ parafermion operators coupled on a triangular lattice, establishing that the Fibonacci phase persists from the quasi-1D limit all the way to the isotropic point and beyond.\cite{MilesDMRG}  A similar state can, interestingly, also emerge from a local bosonic model.\cite{QiSlaveGenon}

On a conceptual level the results reviewed in this section extend the pioneering insights provided by Read and Green\cite{ReadGreen} into more exotic territory.  Indeed the superconducting `Fibonacci phase' bears the same relation to the $\mathbb{Z}_3$ Read-Rezayi state as a spinless $p+ip$ superconductor bears to Moore-Read.  The blueprint of Fig.~\ref{FibFigure} additionally demonstrates proof of concept that well-understood ingredients can combine to yield hardware for a universal topological quantum computer, which is interesting in itself.  An important future challenge involves distilling the architectures discussed  to simpler, more practical forms.  One suggestion invoking remarkably simple quantum Hall bilayers appears in Ref.~\onlinecite{VaeziBarkeshli}.  More generally, we expect that condensing bosons built from fractionalized excitations as done here enables realistic design of many other exotic phases of matter (which frequently appear in toy models featuring multi-particle interactions).

\section{Outlook}

Bloch's band theory provides a successful recipe for the creation of a zero-temperature insulator---assemble electrons into a periodic potential and fill an integer number of bands.  Bardeen-Cooper-Schrieffer theory similarly reveals the main pathway to building a superconductor---incorporate phonon-mediated attraction between electrons in a metal.  We would have loved opening this paper by analogously specifying the ingredients in a tried-and-true recipe for creating a system with non-Abelian anyons and topological ground-state degeneracy.  While this remains a major open question, recent trends suggest that clarity in this direction may  emerge within the next several years following intertwined efforts from the theoretical and experimental community.

Theorists working on the field steadily progress towards a viable recipe by propagating along several complementary trajectories (simultaneously, of course; we do believe in quantum mechanics). One direction---employed extensively in the study of lattice models, non-Abelian quantum Hall states, and many other contexts---starts from exactly solvable Hamiltonians that give rise to a non-Abelian topological phase and proceeds to study its stability.  These solvable models almost certainly do not faithfully represent the microscopics of any solid-state medium.  Nevertheless, given the prized insensitivity of topological phases to microscopic details, one hopes that they are `close enough' to describe the universal behavior of a physical system.  The Moore-Read state, which likely finds realization in GaAs at filling $\nu = 5/2$, provides a notable success of this approach.\cite{Storni}

Another direction, which we reviewed here, attempts to combine reasonably well-understood building blocks (e.g., 1D wires, superconductors, `simple' fractional quantum Hall fluids, etc.)~in exactly the right way to corner the electrons both into a non-Abelian topological phase \emph{and} into the upper-right corner of Fig.~\ref{FisherPlot}.  Both theoretical and experimental considerations motivate this strategy.  On the theory end, it allows us to build a bridge from physics for which we have well-developed theoretical tools into uncharted territories wherein electron-electron correlations produce new phenomena much more exotic than that present in the individual components.  Experimentally, one may argue that we follow the very philosophy of engineering---using naturally available building blocks to create devices that Nature neglects to synthesize for us.  Indeed the roadmap outlined here endeavors to `out-engineer Nature' in a sense, with the goal of designing and controlling novel states of matter that appear difficult to capture intrinsically in isolated materials.

How far we can go in this quest remains to be answered.  The recent wave of experimental activity in the Majorana direction encourages us to continue studying designer non-Abelian platforms, both of the Majorana type and of the more exotic variety---even though the latter's distance from experimental realization should by no means be underestimated.  Aside from fabrication many important challenges exist that we did not review here.  These include devising experimental litmus tests for non-Abelian phases that clearly distinguish from trivial states and establishing manipulation protocols for quantum information applications.  Remarkably, recent years have demonstrated that this study enormously benefits from the ongoing dialogue among high-altitude theorists and down-to-earth experimentalists. The meeting point between these two communities repeatedly proves a very exciting place to dwell, and this case is no exception.

\acknowledgments{
We are indebted to all of our collaborators on work related to non-Abelian statistics, particularly from Refs.~\onlinecite{AliceaBraiding,LindnerParafendleyons,ClarkeParafendleyons,Mong} on which much of this article is based.
J.~A.~gratefully acknowledges funding from the NSF through grant DMR-1341822; the Alfred P.\ Sloan Foundation; the Caltech Institute for Quantum Information and Matter, an NSF Physics Frontiers Center with support of the Gordon and Betty Moore Foundation; and the Walter Burke Institute for Theoretical Physics at Caltech.  A.~S.~gratefully acknowledges support from Microsoft's Station Q, the European Research Council under the European Union's Seventh Framework Programme (FP7/2007-2013) / ERC Project MUNATOP, the US-Israel Binational Science Foundation and  the Minerva Foundation. 
}

\hbadness 10000	
\bibliography{NobelProceedings}

\end{document}